\input{aipcheck}

\documentclass[
    ,final            
  ]
  {aipproc}
\usepackage{amsmath,amssymb}
\layoutstyle{8x11single}

\begin{document}

\title{Directional Search for Isospin-Violating Dark Matter with Nuclear Emulsion}


\author{Keiko I. Nagao}{
  address={KEK Theory Center, IPNS, KEK, 1-1 Oho, Tsukuba, 305-0801, Japan}
}

\author{Tatsuhiro Naka}{
  address={Institute for Advanced Reseach, Nagoya University, Nagoya 464-8602, Japan}
}


\begin{abstract}
Some of direct dark matter searches reported not only positive signals
but also annual modulation of the signal event. However, the parameter
spaces have been excluded by other experiments. Isospin violating dark matter
solves the contradiction by supposing different coupling to proton and neutron.
We study the possibility to test the favored parameter region by isospin 
violating dark matter model with the future detector of dark matter using the nuclear
emulsion. Since the nuclear emulsion detector has directional sensitivity, 
the detector is expected to examine whether the annual modulations observed other experiments 
is caused by dark matter or background signals.
\footnotetext[1]{Talk was presented by K.~I.~Nagao and based on work [1].}
\end{abstract}

\maketitle

\section{Introduction}
Dark matter comes to be known as matter that consists about 23\% of the Universe.
Many experiments have been performed to detect it in direct and indirect ways. 
Some of the direct detections, DAMA, CoGenT and CRESSTII presented data which can be 
interpreted as dark matter signal. Furthermore, they reported the annual modulation of the signal events 
which is expected by the yearly round of the earth.
However the signal parameter regions have been excluded by other late experiments such as XENON 10, XENON 100 and CDMSII.
The discrepancy have received attentions in astro-particle physics.
Isospin violating dark matter  is one of proposals to solve the discrepancy by supposing different dark matter-proton coupling $f_p$ and dark matter-neutron coupling $f_n$ [2]. 
Especially if $f_n/f_p=-0.7$, the signal region of DAMA overlaps that of CoGenT. Besides, part of the region that satisfies both of them is allowed by other null constraints at $m_{\mathrm{DM}}\sim 8$ GeV and $10^{-1}$ pb $\lesssim \sigma_{\mathrm{SI}} \lesssim 10^{-2}$ pb where $\sigma_{\mathrm{SI}}$ is the spin-independent cross section of dark matter-nucleus scattering. 

In this work we examine the possibility that a future direct search project using nuclear emulsion tests the favored region by isospin-violating dark matter. The experiment, which is still in research and development, aims to detect charged particles produced by the dark matter-nucleus scattering. The nuclear emulsion has been employed in experiments to detect charged particles, OPERA project [3] for recent example. It is a kind of photographic film, on which charged particles leave tracks. After development, the tracks become distinguishable on the emulsion layers. The nuclear emulsion detector has the directional sensitivity since the tracks reflect the arrival direction of dark matter. 
Therefore, it can examine whether signal tracks come from the direction expected by the orbital motion of the Earth.
One of the interesting possibility of the experiment is to test the parameter spaces where other experiments observed the annual modulation signal. 
Large mass as solid is another merit of the nuclear emulsion. 
Since solid contains much more number of target nuclei than that of gas detector,  good sensitivity can be  achieved.
It is worth studying the possibility that the direct search with the nuclear emulsion reaches the favored region allowed by both positive and null result experiments in the isospin violating dark matter scenario.

\section{Dark Matter search with nuclear emulsion}
\subsection{Sensitivity for isospin-conserving case}
\begin{table}[hbp]
\begin{minipage}{.6\textwidth}
\begin{center}
\begin{tabular}{|c||r|r|r|}
\hline
\multicolumn{1}{|c|}{\,} & \multicolumn{1}{c|}{Weight(\%)} & \multicolumn{2}{c|}{$A_i$(abundance)} \\ \hline
Ag                                & 39.65                                          &107(51.84) &109(48.16)    \\ 
Br                                 & 29.01                                          &79(50.69)  &81(49.31)      \\ 
O                                  & 11.76                                          &16&                     \\ 
N                                  & 4.57                                            &14&                   \\ 
C                                  & 11.72                                          &12(98.9)    &13(1.1)          \\ \hline
\end{tabular}
\end{center}
\end{minipage}\\
\caption{Contents of nuclear emulsion layers. Mass number of isotope $A_i$ (its natural abundance ratio if it consists more than 1\% of the atom) is shown in the second and third columns. }
\label{Table:isotopes}
\end{table}
The event rate of dark matter detection is defined as 
\begin{eqnarray}
R=N_T n_\chi \int_{E_{R,\mathrm{min}}} dE_R \int_{v_{\mathrm{min}}}^{v_{\mathrm{max}}} d^3 v \,\,f(v) v \frac{d\sigma_A}{dE_R}
\label{eq:eventrate}
\end{eqnarray}
where $N_T$ is the number of target nuclei, $n_\chi$ is the number density of dark matter, $f(v)$ is the distribution function of dark matter velocities,
and $\sigma_A$ is the cross section of the dark matter-nucleus A scattering. 
In this study, sensitivity of the detector is calculated by Monte-Carlo simulation assuming the Maxwell-Boltzmann distribution:
\begin{eqnarray} 
f(v)=\frac{1}{(\pi v_0^2)^{3/2}}e^{-(v+v_E)^2/v_0^2},
\end{eqnarray}
 $v_0=220$ km/sec as the velocity of the Solar-system relative to the galactic halo, $v_E=230$ km/sec as the Earth's velocity relative to the dark matter distribution, and local dark matter density $\rho_0= 0.3$ GeV$/$cm$^{3}$. 
The differential cross section of spin-independent scattering can be represented as $d\sigma_A/dE_R=\sigma_A m_A/(2v^2 \mu_A^2)$ where $m_A$ is the nucleus mass and  $\mu_A$ is the reduced mass, i.e., $\mu_A=m_Am_\chi/(m_A+m_\chi)$.
If there is one kind of atom which one isotope dominates as a target,  the dark matter-nucleus scattering cross section is described as
\begin{eqnarray}
{\sigma}_A = \frac{\mu_A^2}{\Lambda^4}\left[f_p Z F_A^p(E_R) +f_n(A-Z) F_A^n(E_R)\right]^2 
\label{eq:sigmaA}
\end{eqnarray}
where $\Lambda$, $Z$ and $F_A^{p(n)}(E_R)$ are 
a scale which parametrizes the scattering, the atomic number of $A$, and the proton (neutron) form factor, respectively. 
We set both of the form factors are same as $F_A(E_R)$ afterward.
The dark matter-proton coupling is usually  assumed to be same as the dark matter-neutron coupling, i.e. $f_n/f_p=1$. 
We introduce the dark matter-proton scattering cross section as $\sigma_p=f_p^2\mu_p^2F_A(E_R)^2/\Lambda^4$, then Eq.(\ref{eq:sigmaA}) can be written as $\sigma_A=\sigma_p \mu_{A}^2/\mu_p^2 \left[Z +f_n/f_p(A-Z)\right]^2$. 
If there are several isotopes, it yields 
\begin{eqnarray}
\sigma_A=\sigma_p \sum_i \eta_i  \mu_{A_i}^2/\mu_p^2 \left[Z +f_n/f_p(A_i-Z)\right]^2
\end{eqnarray}
where suffixes $i$ labels isotope, and $\eta_i$ is the natural abundance ratio of the isotope.

If there are several target atoms, we should start again from the event rate to derive $\sigma_p$ because parameters other than $\sigma_A$ depend on target atoms.
The number of target nuclei $N_T$ in unit of $g^{-1}$ is represented as $N_0 /\tilde{A}$ where $N_0$ is the Avogadro number and $\tilde{A}$ is the Molar mass in $g/mol$ unit, which is nothing but the atomic mass $A$ in the notation. 
The dark matter-proton cross section is determined by the event rate in unit of /kg/year: 
\begin{eqnarray}
R&=&\sigma_p \sum_j \left[ \xi_j \left(\sum_i \frac{N_0\times 10^3}{A_i^j}  \eta_i \frac{m_{A^j_i}^2}{\mu_p^2} [Z_j+f_n/f_p(A^j_i-Z_j)]^2\right)\right. \nonumber\\
&&\left. \times \left(n_\chi \int_{E_{R,\mathrm{min}}} dE_R \int_{v_{\mathrm{min}}}^{v_{\mathrm{max}}} d^3 v \,\,f(v) v F_{A^j_i}(E_R)^2 \right)\right],
\label{eq:eventrate2}
\end{eqnarray}
where $\xi_j$ is the weight ratio of atom $A^j$ in target matter.

We show the expected 90\% C.L. sensitivity of the nuclear emulsion for isospin-conserving dark matter, i.e. $f_n/f_p=1$, in Figure 1. 
The assumed exposure is 1000 kg$\cdot$year.
In the calculation, perfect detection efficiency and back ground rejection are supposed. 
The target atoms in the nuclear emulsion are listed in Table 1. Silver (Ag) and bromine (Br) are main components as photographic film, and can be target of dark matter scattering. Small grains of AgBr are combined by gelatin, which is mainly consisted of calcium (C), nitrogen (N) and oxygen (O). They also can be target atoms. Only if the light targets are included, the detector has sensitivity for light dark matter with $m_{\mathrm{DM}}\sim 10$ GeV. 
Current limit of the detectable track range is 100 nm on the nuclear emulsion, which corresponds to the energy threshold 160 keV for heavy target atoms, and 33 keV for light ones. 
  
\begin{figure}[t]
\label{fig:emulsionsensitivity}
\includegraphics[scale=.9]{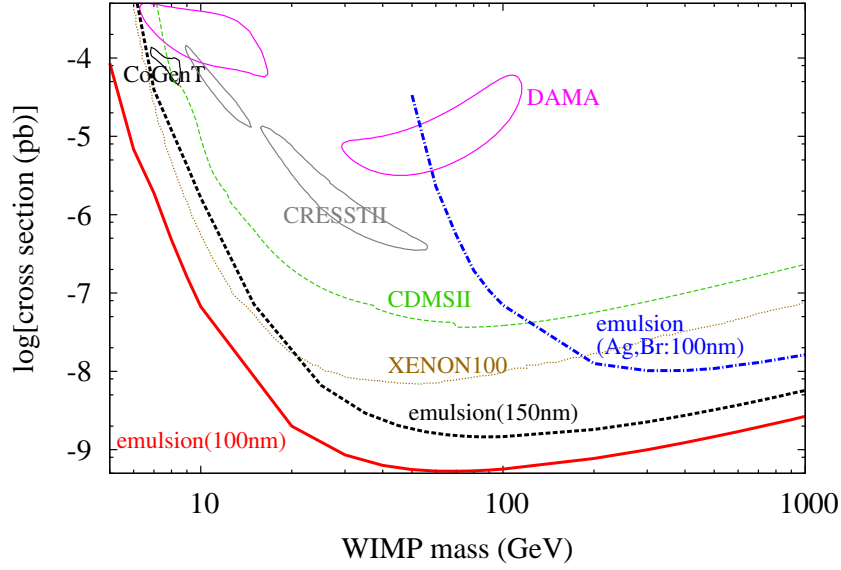}
\caption{Expected sensitivity of the nuclear emulsion detector and current results of direct searches for the isospin-conserving case, i.e. $f_n/f_p=1$. Circles labeled with DAMA, CoGenT and CRESSTII represent the signal regions, while other lines show the constraints or expected sensitivity of the experiments.}
\end{figure}

\subsection{Sensitivity for isospin-violating case}
In this subsection, we convert the sensitivity of the nuclear emulsion shown in Figure 1 to that of the case for $f_n/f_p=-0.7$.
We focus on light mass region since $m_{\mathrm{DM}}\lesssim 10$ GeV is favored by the isospin-violating dark matter. Not heavy target atoms (Ag and Br) but only light ones (C, N, and O) are sensitive to the light dark matter. The light ones have almost same form factors and energy thresholds $\sim 33$ GeV.
In that case, the second parenthesis of Eq.(\ref{eq:eventrate2}) is independent of $i$ and $j$. 
We denote the dark matter-nucleus cross section for $f_n/f_p=1$ as $\sigma_N$, that is nothing but the cross section constrained by experiments.
The ratio of $\sigma_N$ and $\sigma_p$ is derived as
\begin{eqnarray}
\frac{\sigma_p}{\sigma_N}=\frac{\sum_j \xi_j \left(\sum_i \eta_i m_{A_i^j} A_i^{j} \right)}{\sum_j \xi_j \left(\sum_i \eta_i m_{A_i^j}/A_i^j \left[Z_j+f_n/f_p(A_i^j-Z_j)\right]^2\right)} .
\label{eq:ratio_severalatoms}
\end{eqnarray} 

We note that the choice $f_n/f_p=-0.7$ is destructive, especially for atoms of large atomic number. Since such atoms tend to have more number of neutrons than that of protons, the sensitivity suppression is amplified. 
In the case of the nuclear emulsion, light target atoms have small atomic numbers, i.e. the number of neutrons and protons are almost same. It means the sensitivity of the nuclear emulsion detector for isospin-violating dark matter is not so suppressed. 
We show the numerical result in Figure 2. 
Most of signal regions and null constraints are gathered to the cross section $\sigma_p\sim 10^{-2}-10^{-1}$ pb.
On the other hand, sensitivity of the nuclear emulsion is $\sigma_p\sim 10^{-4} -10^{-5}$ pb around $m_{\mathrm{DM}}\sim 8$ GeV, which is triple-digit higher than the favored region.

The favored region is in small mass parameter space near the detector threshold. Therefore, rejection of back ground signals is important. 
Since radioactive backgrounds are much lighter than dark matter particles, 
energy deposit per unit path length dE/dx of the radioactive backgrounds tends to be smaller than that of dark matter signals. 
For example, expected total energy deposition for heavy targets (Ag, Br) and light targets (C, N, O) are
about 1000-2000 keV/$\mu$m and 100-300 keV/$\mu$m, respectively. 
On the other hand, the energy deposit for electron and proton background are about 10 keV/$\mu$m and 50 keV/$\mu$m, respectively. 
By adjusting  the sensitivity of the nuclear emulsion, and by the sensitivity control in development process, producing nuclear emulsion which does not have sensitivity to the background is expected to be possible. 

\begin{figure}[t]
\label{fig:emulsionsensitivity}
 \begin{minipage}{0.7\hsize}
\includegraphics[scale=.9]{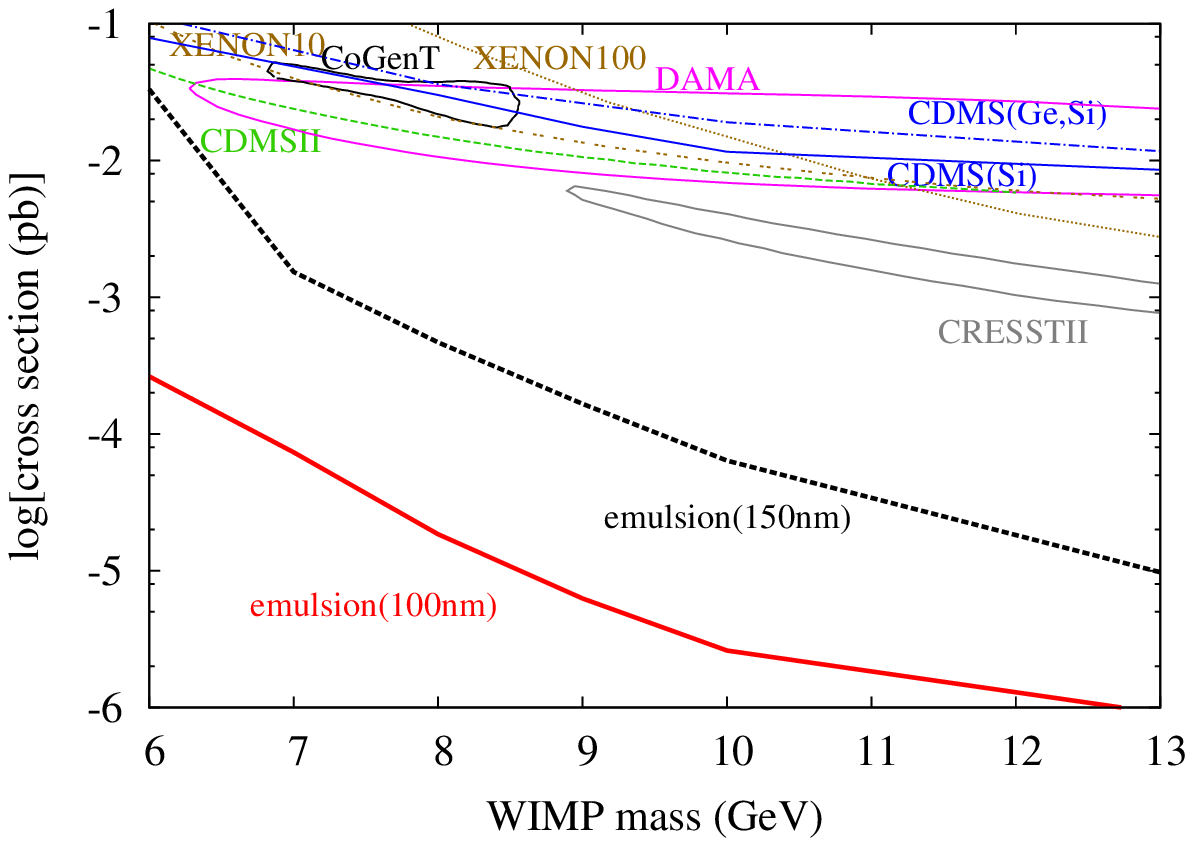}
 \end{minipage}
 \begin{minipage}{0.29\hsize}
 \hspace{-2.em}
   \includegraphics[scale=.85]{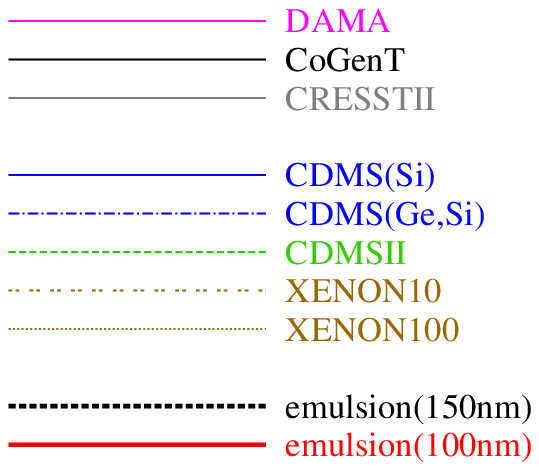}
 \end{minipage}
 \caption{Expected sensitivity of the nuclear emulsion detector to isospin-violating dark matter. The violation $f_n/f_p=-0.7$ is supposed.}
\end{figure}

\section{Conclusion}
We study the possibility that the dark matter search project using the nuclear emulsion can test the parameter space favored by isospin-violating dark matter. 
Since the detector have directional detectability, it can focus on signals come from direction of dark matter wind.
Therefore, examining the signal regions of other experiments by the nuclear emulsion is one of interesting possibilities of the detector.
The nuclear emulsion is expected to have enough sensitivity to test the favored region, i.e., $\sigma_p=10^{-1}-10^{-2}$ pb and $m_{\mathrm{DM}}\sim 8$ GeV, if good detection efficiency and back ground rejection are achieved. 




\section*{References}
[1] K.~I.~Nagao and T.~Naka,
  arXiv:1205.0198 [hep-ph].
  
\noindent [2]  J.~L.~Feng, J.~Kumar, D.~Marfatia and D.~Sanford,
  Phys. Lett.  B {\bf 703}, 124 (2011)
  [arXiv:1102.4331 [hep-ph]].

\noindent [3]  N.~Agafonova {\it et al.}  [OPERA Collaboration],
  Phys.\ Lett.\  B {\bf 691}, 138 (2010)
  [arXiv:1006.1623 [hep-ex]].
\end{document}